\documentclass[12pt]{article}
\usepackage[english,german,french,polish]{babel}
\usepackage[T1]{fontenc}
\usepackage{amsfonts}

\begin{document}

\selectlanguage{english}

\baselineskip 0.73cm
\topmargin -0.4in
\oddsidemargin -0.1in

\let\ni=\noindent

\renewcommand{\thefootnote}{\fnsymbol{footnote}}

\newcommand{\SM}{Standard Model }

\newcommand{\SMo}{Standard-Model }

\pagestyle {plain}

\setcounter{page}{1}



~~~~~~
\pagestyle{empty}

\begin{flushright}
IFT-- 11/~1
\end{flushright}

\vspace{1.0cm}

{\large\centerline{\bf Hidden-sector correction to Coulomb potential}}

{\large\centerline{\bf through the photonic portal{\footnote{Work supported in part by Polish MNiSzW scientific research grant N N202 103838 (2010--2012).}} }}

\vspace{0.5cm}

{\centerline {\sc Wojciech Kr\'{o}likowski}}

\vspace{0.3cm}

{\centerline {\it Institute of Theoretical Physics,  Faculty of Physics, University of Warsaw}}

{\centerline {\it Ho\.{z}a 69, 00--681 Warszawa, ~Poland}}

\vspace{0.6cm}

{\centerline{\bf Abstract}}

\vspace{0.2cm}

We show that in the model of hidden sector of the Universe, interacting with the \SMo sector through the photonic portal, the \SMo Coulomb potential gets a tiny hidden-sector additive correction that might turn out to be either exciting or fatal for the verification of this model.
 
\vspace{0.6cm}

\ni PACS numbers: 14.80.-j , 04.50.+h , 95.35.+d 

\vspace{0.6cm}

\ni January  2011

\vfill\eject

\pagestyle {plain}

\setcounter{page}{1}

\vspace{0.3cm}

\ni {\bf 1. Introduction}

\vspace{0.3cm} 

A model of hidden sector of the Universe, proposed in previous papers [1,2], consists of sterile spin-1/2 Dirac  fermions ("\,$\!$sterinos"), sterile spin-0 bosons ("\,$\!$sterons") and sterile nongauge mediating bosons ("$A$ bosons") described by an antisymmetric-tensor field (of dimension one) weakly coupled to steron-photon pairs as well as to antisterino-sterino pairs, giving the new weak interaction

\begin{equation}
- \frac{1}{2} \sqrt{\!f\,}\left(\varphi F_{\mu \nu} + \zeta \bar\psi \sigma_{\mu \nu} \psi \right) A^{\mu \nu}\,.
\end{equation}

\ni Here, $F_{\mu \nu} = \partial_\mu A_\nu - \partial_\nu A_\mu $ is the \SMo electromagnetic field (of dimension two), while $\sqrt{\!f\,}$ and $\sqrt{\!f\,}\zeta$ denote two dimensionless small coupling constants. We  presume that  

\begin{equation}
\varphi = <\!\!\varphi\!\!>_{\rm vac}\! + \,\varphi_{\rm ph}
\end{equation}
 
\ni with a spontaneously nonzero vacuum expectation value $<\!\!\varphi\!\!>_{\rm vac}\, \neq 0$. We have called such a coupling of photons to the hidden sector "photonic portal"\, (to the hidden sector), and considered it as an alternative to the popular "Higgs portal" \,[3].

In the present note, we show that the first term in the interaction (1) implies a small correction to the \SMo Coulomb potential. It can be estimated, if the coupling constant $f$, the vacuum expectation value $<\!\!\varphi\!\!>_{\rm vac}$ and the masses of sterino, steron and $A$ boson, $m_\psi, m_\varphi$ and $M$, are somehow established.

The new interaction Lagrangian (1), jointly with the $A$-boson kinetic and \SMo electromagnetic Lagrangians, leads to the following field equations for  $F_{\mu \nu}\!$ and  $A_{\mu \nu}$:

\vspace{-0.3cm} 

\begin{equation}
\partial^\nu\!\! \left[F_{\mu \nu} +  \sqrt{\!f\,}(<\!\!\varphi\!\!>_{\rm vac}\! + \,\varphi_{\rm ph}) A_{\mu \nu}\right] = -j_\mu \;\;,\;\; F_{\mu \nu} = \partial_\mu A_\nu - \partial_\nu A_\mu 
\end{equation}

\ni and

\vspace{-0.3cm} 

\begin{equation}
(\Box - M^2)A_{\mu \nu} = - \sqrt{\!f\,}\left[ (<\!\!\varphi\!\!>_{\rm vac}\! + \,\varphi_{\rm ph}) F_{\mu \nu} + \zeta \bar\psi \sigma_{\mu \nu} \psi\right]\,, 
\end{equation}

\ni  where $j_\mu$ denotes the \SMo electric current  and $M$ stands for a mass scale of $A$ bosons, expected typically to be large.

The field equations (3) are Maxwell's equations modified in the presence of hidden sector interacting with the \SMo sector through our photonic portal. This modification is of magnetic character, because the hidden-sector contribution to the total electric source-current 

\begin{equation}
j_\mu + \partial^\nu[ \sqrt{\!f\,}(<\!\!\varphi\!\!>_{\rm vac}\! + \,\varphi_{\rm ph}) A_{\mu \nu}]
\end{equation}

\ni of the electromagnetic field $A_\mu $ is a four-divergence giving no contribution to the total electric charge

\begin{equation}
\int d^3x\{j_0 + \partial^k[ \sqrt{\!f\,}(<\!\!\varphi\!\!>_{\rm vac}\! + \,\varphi_{\rm ph}) A_{0 k}]\} = \int d^3x j_0 = Q.
\end{equation}

It can be seen that the vacuum expectation value $<\!\!\varphi\!\!>_{\rm vac}\, \neq 0$ generates spontaneously a small sterino magnetic moment

\begin{equation}
\mu_\psi = \frac{f \zeta}{M^2}<\!\!\varphi\!\!>_{\rm vac} \,, 
\end{equation}

\ni although sterinos are electrically neutral. This is a consequence of an effective sterino magnetic interaction
 
\begin{equation}
- \frac{1}{2}\mu_\psi \bar\psi \sigma_{\mu \nu} \psi F^{\mu \nu}  
\end{equation}

\ni appearing, when the low-momentum-transfer approximation

\begin{equation}
A_{\mu \nu} \simeq \frac{\sqrt{\!f\,}\,\zeta}{M^2}\bar\psi \sigma_{\mu \nu} \psi 
\end{equation}

\ni effectively implied by Eq. (4) with $F_{\mu \nu}\rightarrow 0$ is used in the second term of weak interaction (1). 

It will turn out, however, that due to the photonic portal the \SMo electromagnetic interaction, in particular the Coulomb potential, gets a tiny hidden-sector correction. 

\vspace{0.3cm}

\ni {\bf 2. Hidden-sector correction to Coulomb potential}

\vspace{0.3cm} 

From Eq.  (4) we can see that

\begin{equation}
(\Box - M^2)\partial^\nu A_{\mu\nu} = -\sqrt{\!f\,}\partial^{\,\nu} \left(\varphi F_{\mu \nu} + \zeta \bar\psi \sigma_{\mu \nu} \psi\right)  \,.
\end{equation}

\ni Hence, in the case of $\varphi_{\rm ph} \rightarrow 0$, $\psi \rightarrow 0$ and $A_{\mu \nu} \rightarrow A_{\mu \nu}^{(\rm vac)}$ we get

\begin{equation}
(\Box - M^2)\partial^{\,\nu} A_{\mu \nu}^{(\rm vac)} =  -\sqrt{\!f\,}\!\!<\!\!\varphi\!\!>_{\rm vac}\partial^{\,\nu} F_{\mu \nu} \,, 
\end{equation}

\ni  where from Eq. (3)

\vspace{-0.3cm} 

\begin{equation}
\partial^{\,\nu} F_{\mu \nu}  = - j_\mu - \sqrt{\!f\,}\!\!<\!\!\varphi\!\!>_{\rm vac}\partial^{\,\nu} A_{\mu \nu}^{(\rm vac)} \,. 
\end{equation}

\ni  Thus,

\vspace{-0.3cm} 

\begin{equation}
(\Box - \tilde{M}^2)\partial^{\,\nu} A_{\mu \nu}^{(\rm vac)} =  \sqrt{\!f\,}\!\!<\!\!\varphi\!\!>_{\rm vac}j_\mu 
\end{equation}

\ni  with

\vspace{-0.3cm} 

\begin{equation}
\tilde{M}^2 = M^2 + f <\!\!\varphi\!\!>_{\rm vac}^2\,. 
\end{equation}

\ni For a \SMo point-charge at rest at $\vec{x}_0$ we have 
 
\begin{equation}
j_\mu (\vec{x}) =  g_{\mu 0} e_0 \delta^3(\vec{x} - \vec{x}_0) \,,  
\end{equation}

\ni and then Eq. (13) gives

\begin{equation}
\partial^{\,\nu} A_{\mu \nu}^{(\rm vac)}(x) =  -g_{\mu 0} \sqrt{\!f\,}\!\!<\!\!\varphi\!\!>_{\rm vac} \frac{e_0}{4\pi| \vec{x} - \vec{x}_0|} e^{-\tilde{M}|\vec{x} - \vec{x}_0|}\,. 
\end{equation}

Therefore, a hidden-sector additive correction to electrostatic energy of two point-charges at rest at $\vec{x_0}$ and $\vec{\,x}'_0$, following from the first term of weak interaction (1), arises and is equal to

\begin{equation} 
\delta V^{(\rm vac)} = -\frac{1}{2}\!\int d^3x \sqrt{\!f\,}\!\!<\!\!\varphi\!\!>_{\rm vac} F^{\mu \nu}(x)' A_{\mu \nu}^{(\rm vac)}(x) = -\int\! d^3x \sqrt{\!f\,}\!\!<\!\!\varphi\!\!>_{\rm vac} A^\mu(x)' \partial^\nu\! A^{(\rm vac)}_{\mu \nu}(x)\,,
\end{equation}

\ni where $F^{\mu \nu}(x)' = \partial^\mu A^\nu(x)' - \partial^\nu A^\mu(x)'$ and

\begin{equation} 
A^\mu(x)'= g^{\mu 0} \left[\frac{e'_0}{4\pi| \vec{x} - \vec{\,x}'_0|} + O(f)\right] \,, 
\end{equation}

\ni the latter is due to Eqs. (12) (with $\partial_\nu A^\nu(x)' = 0$) and (15) (with $e'_0$ and $\vec{\,x}'_0$). This implies together with Eq. (16) that

\begin{eqnarray}
\delta V^{(\rm vac)} & = & f\!<\!\!\varphi\!\!>^2_{\rm vac}\!\int\!\! d^3x \frac{e_0\,e'_0}{(4\pi)^2|\vec{x} -\vec{x}_0||\vec{x} - \vec{\,x}'_0|} e^{-\tilde{M}|\vec{x} - \vec{x}_0|} + O(f^2) \nonumber \\
& = & \frac{f\!<\!\!\varphi\!\!>^2_{\rm vac}\!}{\tilde{M}^2}\frac{e_0\,e'_0}{4\pi|\vec{x}_0 -\vec{\,x}'_0|} \left(1 - e^{-\tilde{M}|\vec{x}_0 - \vec{\,x}'_0|}\right) + O(f^2) \,.
\end{eqnarray}

We can see that such a hidden-sector correction $\delta V^{(\rm vac)}$ to the Coulomb potential $V = e_0\,e'_0/(4\pi|\vec{x}_0 -\vec{\,x}'_0|)$ is smaller than the latter by the factor

\vspace{-0.2cm}

\begin{equation} 
\frac{ f\!<\!\!\varphi\!\!>^2_{\rm vac}}{\tilde{M}^2}\left(1 - e^{-\tilde{M}|\vec{x}_0 - \vec{\,x}'_0|}\right)\,,
\end{equation} 

\ni where $f\!<\!\!\varphi\!\!>^2_{\rm vac}/\tilde{M}^2$ is expected to be tiny. This factor approaches $f\!<\!\!\varphi\!\!>^2_{\rm vac} |\vec{x}_0 -\vec{\,x}'_0|/\tilde{M}$ or $f\!<\!\!\varphi\!\!>^2_{\rm vac}/\tilde{M}^2$ if $\tilde{M}|\vec{x}_0 -\vec{\,x}'_0|$ tends to 0 or $\infty$, respectively.

\vspace{0.3cm}

\ni {\bf 3. A suggestion for sterile masses consistent with thermal sterinos}

\vspace{0.3cm}

In our model of hidden sector, thermal sterinos are candidates for cold dark matter after their freeze out. Thus, with the use of the abundance of cold dark matter observed by WMAP collaboration, $\Omega_{\rm DM} h^2 \simeq 0.11$ [4], we infer that in the case of our weak interaction (1) the thermal average of total annihilation cross-section of an antisterino-sterino pair, multiplied by the sterino relative velocity, can be estimated as  

\begin{equation}
<\!\!\sigma_{\rm ann} (\bar{\psi} \psi) 2v_\psi\!\!> \simeq {\rm p b} \simeq \frac{8}{\pi}\, \frac{10^{-3}}{\rm TeV^2}\,,
\end{equation}

\ni where 

\begin{equation}
\sigma_{\rm ann} (\bar{\psi} \psi) 2v_\psi \sim \left[\sigma (\bar{\psi} \psi \rightarrow \gamma\varphi_{\rm ph}) + \sum_f \sigma (\bar{\psi} \psi \rightarrow \bar{f} f)\right] 2 v_\psi
\end{equation}

\vspace{-0.2cm}

\ni with $f$'s denoting charged leptons $e^-, \mu^-, \tau^-$ as well as quarks $u, c, t$ and $d, s, b$ (among quarks the top $t$ is considered or omittted if $m_t < m_\psi$ or $m_t > m_\psi$, respectively). Here, the interaction (1) plus the \SMo electromagnetic coupling $-j_\mu A^\mu$ gives ({\it cf.} the third Ref. [1] and the second Ref. [2]):

\begin{equation}
\sigma(\bar{\psi} \psi \rightarrow \gamma\varphi_{\rm ph}) 2v_\psi  = \frac{1}{6\pi} \left(\frac{f\,\zeta}{M^2}\right)^2 (E^2_\psi + 2 m^2_\psi)\left(1- \frac{m^2_\varphi}{4E^2_\psi} \right)
\end{equation}

\vspace{-0.2cm}

\ni and 

\vspace{-0.2cm}

\begin{equation}
\sigma (\bar{\psi} \psi \rightarrow \bar{f} f) 2v_\psi = \frac{1}{12\pi} \left(\frac{e_f f\,\zeta\!<\!\!\varphi\!\!>_{\rm vac}}{M^2}\right)^{\!\!2}  \frac{E^2_\psi + 2 m^2_\psi}{E^2_\psi} \,,
\end{equation}

\ni where the fermion mass $m_f$ is neglected {\it versus} sterino mass $m_\psi$. The annihilation processes $
\bar{\psi} \psi \rightarrow \gamma\varphi_{\rm ph}$ and  $\bar{\psi} \psi \rightarrow \bar{f} f$ go virtually as $ \bar{\psi} \psi \rightarrow A^*\!\rightarrow \gamma\varphi_{\rm ph}$ and $\bar{\psi} \psi \rightarrow A^*\!\rightarrow \gamma^*\!\rightarrow \bar{f} f$, respectively, where $A^*\!\rightarrow \gamma^*$ involves the action of $<\!\!\varphi\!\!>_{\rm vac}$. Thus,

\begin{equation}
\sigma_{\rm ann}(\bar{\psi} \psi) 2v_\psi \!\sim\!\left[\!1\!+\!\left(\!\frac{20}{3}\,{\rm for}\,m_\psi \!<\! m_t\:\, {\rm or}\:\, 8\,{\rm for}\, m_\psi \!>\! m_t \!\right)\frac{\sigma(\bar{\psi} \psi \!\rightarrow\! e^+ e^-)}{\sigma(\bar{\psi} \psi \!\rightarrow\! \gamma\varphi_{\rm ph})}\!\right]\sigma(\bar{\psi} \psi \!\rightarrow\! \gamma\varphi_{\rm ph})2v_\psi \,.
\end{equation}

\ni Here, $3+3\cdot 3(4/9 + 1/9) = 8$ for $m_\psi>m_t$ and $3+2\cdot 3(4/9 + 1/9) + 3(1/9) = 20/3$ for $m_\psi<m_t$.

Now, we assume tentatively that

\vspace{-0.2cm}

\begin{equation}
m_\psi^2 \sim m_\varphi^2 \sim (10^{-2} \;{\rm to}\; 1) M^2 \sim \,<\!\!\varphi\!\!>_{\rm vac}^2
\end{equation}

\ni and conjecture boldly that

\vspace{-0.3cm}

\begin{equation}
f \sim e^2 = 4\pi \alpha \simeq 0.0917 \;\;\;,\;\;\; \zeta \sim 1 \,.
\end{equation}

\ni Then, from Eqs. (25) and (26)

\vspace{-0.3cm}

\begin{equation}
\sigma_{\rm ann}(\bar{\psi} \psi) 2v_\psi \!\sim\!\left[1+\left(\frac{20}{3}\,{\rm for}\,m_\psi \!<\! m_t\: {\rm or}\: 8\,{\rm for}\, m_\psi \!>\! m_t \right)\frac{2}{3}e^2\right] \frac{3f^2}{8\pi m_\psi^2}\times(10^{-4}\,{\rm to}\, 1)
\end{equation}

\ni with $f^2 = e^4$ due to Eq. (27), where $E_\psi \sim m_\psi$ and the electron mass $m_e$ is neglected. Note that due to Eq. (28) $<\!\sigma_{\rm ann}(\bar{\psi} \psi)2v_\psi\! > \,\sim \sigma_{\rm ann}(\bar{\psi} \psi) 2v_\psi $.

From Eqs. (21) and (28) with $f^2 = e^4$ we estimate

\begin{equation}
m_{\psi} \sim \,(7.5 \;\;{\rm to}\;\; 770) \;{\rm GeV} 
\end{equation}

\ni and

\vspace{-0.3cm}

\begin{equation}
M \sim (10\;{\rm to}\;1) m_\psi \sim (75\;{\rm to}\; 770)\;{\rm GeV} \,.
\end{equation}

In particular, for $M \sim 10m_\psi \sim 75$ GeV the numerical coefficient in the factor (20) becomes

\begin{equation}
\frac{ f\!<\!\!\varphi\!\!>^2_{\rm vac}}{\tilde{M}^2} =  \frac{ f\!<\!\!\varphi\!\!>^{\!2}_{\rm vac}/M^2}{1+ f\!<\!\!\varphi\!\!>^{\!2}_{\rm vac}/M^2} \sim \frac{e^2\, m^2_\psi/M^2}{1+e^2 m^2_\psi M^2}\sim 0.00092\,.
\end{equation}

\ni Of course, this estimate is simply a guess, but it is an example consistent with the thermal sterinos playing the role of cold dark matter after their freeze-out. For a smaller value of $f$ than $f\sim e^2$, the coefficient (31) is decreased proportionally to $f$ when $f^2/m_\psi^2$ (due to Eqs. (21) and (28)) as well as $m^2_\psi/M^2$ (due to Eqs. (26)) are fixed.

The not observed yet hidden-sector additive correction to Coulomb potential, generated through the photonic portal, might turn out to be either exciting or fatal for the verification of our model of the hidden sector of the Universe.


\vspace{1.5cm}

{\centerline{\bf References}}

\vspace{0.4cm}

\baselineskip 0.73cm

{\everypar={\hangindent=0.65truecm}
\parindent=0pt\frenchspacing

{\everypar={\hangindent=0.65truecm}
\parindent=0pt\frenchspacing

[1]~W.~Kr\'{o}likowski,{\it Acta Phys. Polon.} {\bf B 39}, 1881 (2008); arXiv: 0803.2977 [{\tt hep--ph}]; {\it Acta Phys. Polon.} {\bf B 40}, 111 (2009); {\it Acta Phys. Polon.} {\bf B 40}, 2767 (2009). 

\vspace{0.2cm}

[2]~W.~Kr\'{o}likowski,  arXiv: 0911.5614 [{\tt hep--ph}]; {\it Acta Phys. Polon.} {\bf B 41}, 1277 (2010). 

\vspace{0.2cm}

[3]~{\it Cf. e.g.} J. March-Russell, S.M. West, D. Cumberbath and D.~Hooper, {\it J. High Energy Phys.} {\bf 0807}, 058 (2008). 

\vspace{0.2cm}

[4]~K. Nakamura {\it et al.} (Particle Data Group), {\it J. of Phys.}, {\bf G 37}, 075021 (2010). 

\vspace{0.2cm}

\vspace{0.2cm}

\vfill\eject

\end{document}